\documentclass[sigconf]{acmart}

\AtBeginDocument{%
  \providecommand\BibTeX{{%
    \normalfont B\kern-0.5em{\scshape i\kern-0.25em b}\kern-0.8em\TeX}}}

\begin{document}

\title{ContextNet: A Click-Through Rate Prediction Framework Using Contextual information to Refine Feature Embedding}


\author{Zhiqiang Wang, Qingyun She, PengTao Zhang, Junlin Zhang}
\affiliation{
     \institution{Sina Weibo Corp}
     \city{Beijing}
     \country{China}}
\email{roky2813@sina.com,qingyun_she@163.com,{pengtao1,junlin6}@staff.weibo.com}

\begin{abstract}
  Click-through rate (CTR)  estimation is a fundamental task in personalized advertising and recommender systems and it's important for ranking models to effectively capture complex high-order features. Inspired by the success of ELMO and Bert in NLP field, which dynamically refine word embedding according to the context sentence information where the word appears, we think it's also important to dynamically refine each feature's embedding layer by layer according to the context information contained in input instance in CTR estimation tasks. We can effectively capture the useful feature interactions for each feature in this way. In this paper, We propose a novel CTR Framework named ContextNet that implicitly models high-order feature interactions by dynamically refining each feature's embedding according to the input context. Specifically, ContextNet consists of two key components: contextual embedding module and ContextNet block. Contextual embedding module aggregates contextual information for each feature from input instance and ContextNet block maintains each feature's embedding layer by layer and dynamically refines its representation by merging contextual high-order interaction information into feature embedding. To make the framework specific, we also propose two models(ContextNet-PFFN and ContextNet-SFFN) under this framework by introducing linear contextual embedding network and two non-linear mapping sub-network in ContextNet block. We conduct extensive experiments on four real-world datasets and the experiment results demonstrate that our proposed ContextNet-PFFN and ContextNet-SFFN model outperform state-of-the-art models such as DeepFM and xDeepFM significantly.
\end{abstract}

\maketitle

\section{Introduction}
Click-through rate (CTR) estimation has become one of the most essential tasks in many real-world applications. Many models have been proposed to resolve this problem such as Logistic Regression (LR) \cite{10.1145/2487575.2488200}, Polynomial-2 (Poly2) \cite{rendle2010factorization}, tree-based models \cite{he2014practical}, tensor-based models \cite{koren2009matrix}, Bayesian models \cite{graepel2010web}, and Field-aware Factorization Machines (FFMs) \cite{juan2016field}.

Deep learning techniques have shown promising results in many research fields such as computer vision \cite{krizhevsky2012imagenet, he2016deep}, speech recognition \cite{graves2013speech,tang2017end} and natural language understanding \cite{cho2014learning,mikolov2010recurrent}. As a result, employing DNNs for CTR estimation has also been a research trend in this field \cite{zhang2016deep,cheng2016wide,xiao2017attentional,guo2017deepfm,lian2018xdeepfm,qu2016product,wang2017deep}. Some deep learning based models have been introduced and achieved success such as Factorisation-Machine Supported Neural Networks(FNN)\cite{zhang2016deep}, Attentional Factorization Machine (AFM)\cite{cheng2016wide}, wide\&deep\cite{xiao2017attentional}, DeepFM\cite{guo2017deepfm}, xDeepFM \cite{lian2018xdeepfm}, DIN\cite{zhou2018deep} etc.

Feature interaction is critical for CTR tasks and it’s important for these ranking models to effectively capture  complex features. Most DNN ranking models such as FNN and DeepFM use the shallow MLP layers to model high-order interactions in implicit way which has been proved to be ineffective \cite{beutel2018latent}. Some CTR model like xDeepFM \cite{lian2018xdeepfm}  explicitly introduces high-order feature interactions by adding sub-network into the network structure. However, that will  significantly increases the computation time and it's hard to deploy it in real-world application.

Inspired by the success of  ELMO\cite{peters2018deep} and Bert\cite{devlin2018bert} in NLP field, which  dynamically refine word embedding according to the sentence context where the word appears, we think it's also important to dynamically change the feature embedding according to other contextual features in the same instance it appears in CTR tasks. We can effectively capture the useful feature interactions for each feature by introducing the context aware feature embedding into CTR models.

Though AutoInt\cite{song2019autoint} and Fi-GNN\cite{2019Fi} can also dynamically change feature embedding as our proposed model does,  feature representation of these models is kind of weighted summation of  pair-wise interaction . These models follow the rule of feature aggregation in summation way after pair-wise interaction, while our proposed model follow the rule of feature interaction in multiplicative way after feature aggregation by a specific network. Alex Beutel et.al\cite{2018Latent} have proved that  addictive feature interaction is inefficient in capturing common feature crosses. They proposed  a simple but effective  approach named “latent cross”  which is a kind of multiplicative interactions between the context embedding and the neural network hidden states in RNN model. Our work is inspired by  both  the Bert and “latent cross”.

In this work, We propose a new CTR framework named ContextNet which can dynamically refine feature's embedding according to the context it appears and effectively model high-order feature interactions for each feature. Specifically, ContextNet consists of two key components: contextual embedding module and ContextNet block. Contextual embedding module aggregates contextual information for each feature from input instance and ContextNet block maintains each feature's embedding layer by layer and dynamically refines its representation by merging contextual high-order interaction information into feature embedding. So the ContextNet provides a flexible mechanism for each feature to dynamically and efficiently filter out the most useful high-order cross information for its own purpose in current context it appears. Another advantage of ContextNet over most DNN models is that it has good model interpretability. Notice that ContextNet is a new CTR framework instead of a specific model, which means we can design various detailed models based on this framework. We also propose two ContextNet based models in this paper in order to make the ContextNet framework specific and experimental results prove its effectiveness.

The contributions of our work are summarized as follows:

\begin{enumerate}
  \item We propose a novel CTR Framework named ContextNet that implicitly models high-order feature interactions by dynamically refining the feature embedding according to the context information contained in input instance.

  \item To make the ContextNet framework specific, we propose a contextual embedding network and two non-linear mapping sub-network in ContextNet block. So we design two specific ContextNet-based models in our work under the proposed framework, which is named ContextNet-PFFN and ContextNet-SFFN, respectively.

  \item We conduct extensive experiments on four real-world datasets and the experiment results demonstrate that our proposed ContextNet-PFFN and ContextNet-SFFN outperform state-of-the-art models significantly.
\end{enumerate}

The rest of this paper is organized as follows. Section 2 introduces some related works which are relevant with our proposed model. We introduce our proposed ContextNet framework in detail in Section 3. The experimental results on four real world datasets are presented and discussed in Section 4. Section 5 concludes our work in this paper.

\section{Related Work}
\subsection{Context Aware Word Embedding in NLP}
Word embedding is a very important concept in NLP and it attempts to map words from a discrete space into a semantic space. In early stage, Word2vec\cite{mikolov2013efficient} and GloVe\cite{pennington2014glove} learn a constant embedding vector for a word and the embedding is  same for a word in different sentences.  However, it's obvious that polysemous word should have different embedding in various sentence contexts. To deal with this issue, context information of the sentence is used to predict a dynamic word embedding. For example, ELMO\cite{peters2018deep} uses the bidirectional RNN to model the context information. The GPT\cite{radford2018improving} and Bert\cite{devlin2018bert} model leverages the Transformer\cite{vaswani2017attention} model to jointly consider both the left and right context information in the sentence. Our work is inspired by these context aware word embedding approaches and we introduce context aware feature embedding into CTR tasks in this work.

\subsection{Deep Learning based CTR Models}
Many deep learning based CTR models have been proposed in recent years and how to effectively model the feature interactions is the key factor for most of these neural network based models.

Factorization-Machine Supported Neural Networks (FNN)\cite{zhang2016deep} is a feed-forward neural network using FM to pre-train the embedding layer. Wide \& Deep Learning\cite{xiao2017attentional} jointly trains wide linear models and deep neural networks to combine the benefits of memorization and generalization for recommender systems. However, expertise feature engineering is still needed on the input to the wide part of Wide \& Deep model. To alleviate manual efforts in feature engineering, DeepFM\cite{guo2017deepfm} replaces the wide part of Wide \& Deep model with FM and shares the feature embedding between the FM and deep component. Most DNN CTR models rely on  two or three MLP layers to model the high-order interactions in an implicit way and some research\cite{beutel2018latent} has proved MLP is an ineffective way to capture the high-order interactions.

Some works explicitly introduce high-order feature interactions by sub-network. Deep \& Cross Network (DCN)\cite{wang2017deep} efficiently captures feature interactions of bounded degrees in an explicit fashion. Similarly, eXtreme Deep Factorization Machine (xDeepFM) \cite{lian2018xdeepfm} also models the low-order and high-order feature interactions in an explicit way by proposing a novel Compressed Interaction Network (CIN) part.  FiBiNET\cite{huang2019fibinet} can dynamically learn feature importance via the Squeeze-Excitation network (SENET) mechanism and feature interactions via bilinear function. AutoInt\cite{song2019autoint} proposes a multi-head self-attentive neural network with residual connections to explicitly model the feature interactions in the low-dimensional space. Fi-GNN\cite{2019Fi} represents the multi-field features in a graph structure, where each node corresponds to a feature field and different fields can interact through edges. Though AutoInt[25] and Fi-GNN\cite{2019Fi} can also dynamically change feature embedding by proposing a multi-head self-attentive neural network or graph neural network, feature representation of these models is kind of weighted summation of pair-wise interaction. Many research\cite{2018Latent,2020Neural} have proved that addictive feature interaction is inefficient in capturing common feature crosses. Our proposed models collect global contextual information by an independent contextual embedding network to change the feature representation in the multiplicative way. Our experimental results show this advantage.

\section{Our Proposed Model}

In this section, we will firstly introduce ContextNet framework and then describe the key components in detail in the following sections.

\begin{figure*}
  \setlength{\abovecaptionskip}{0pt}
  \centering
  \includegraphics[width=0.85\linewidth]{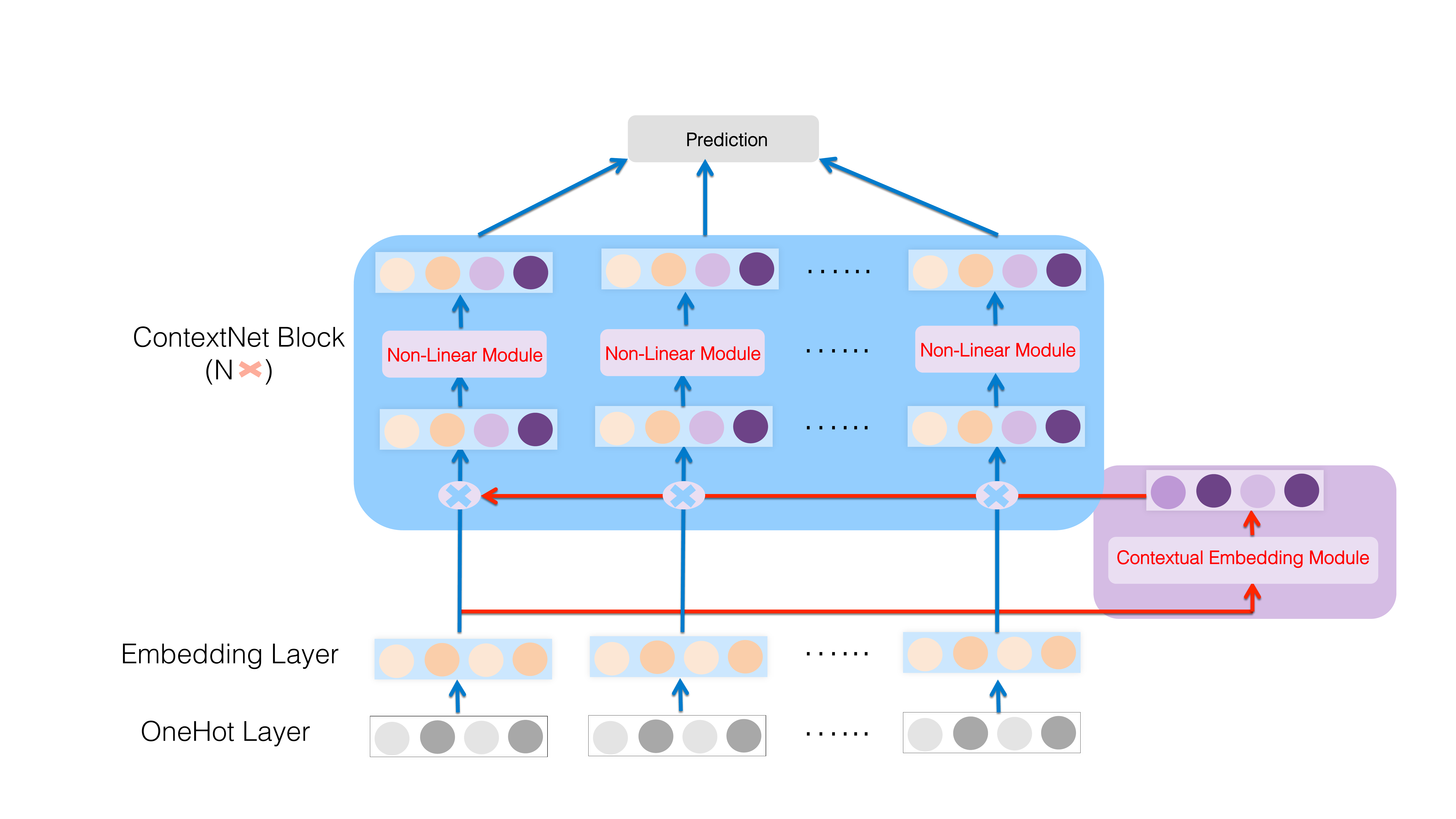}
  \caption{The Neural Structure of ContextNet  Framework}
  \label{fig:contextnet}
\end{figure*}

\subsection{ContextNet Framework}
As depicted in Figure \ref{fig:contextnet},  we propose a novel CTR Framework named ContextNet that implicitly models high-order feature interactions by dynamically refining the feature embedding according to the context information contained in input instance. ContextNet consists of two variable components: contextual embedding module and ContextNet block. Contextual embedding module aggregates contextual information in the same instance for each feature and projects the collected contextual information to the same low-dimensional space as feature embedding lies in. Notice that the input of contextual embedding module is always from the feature embedding layer. ContextNet block implicitly model high-order interactions by merging contextual information into each feature's feature embedding firstly, and then conduct the non-linear transformation on the merged embedding in order to  better capture high-order interactions. We can stack ContextNet  block by block to form deeper network and the refined feature's embedding output of the previous block is the input of the next one. The different ContextNet block has the corresponding contextual embedding module to refine each feature’s embedding. The final ContextNet block’s output is feed into the prediction layer to give the instance's prediction value.


\subsection{Feature Embedding}
The input data of CTR tasks usually consists of sparse and dense features. Such features are  encoded as one-hot vectors which often lead to excessively high-dimensional feature spaces for large vocabularies. The common solution to this problem is to introduce the embedding layer. Generally, the sparse input can be formulated as:
\begin{equation}
  x = [x_1, x_2, ..., x_f]
\end{equation}
where $f$ denotes the number of fields, and $x_i \in \mathbb{R}^n$ denotes a one-hot vector for a categorical  field with $n$ features and $x_i \in \mathbb{R}^n$ is vector with only one value for a numerical  field. We can obtain feature embedding $E_i$ for one-hot vector $x_i$ via:
\begin{equation}
  E_i = W_ex_i
\end{equation}
where $W_e \in \mathbb{R}^{k\times n}$ is the embedding matrix of $n$ features and $k$ is the dimension of field embedding. The numerical feature $x_j$ can also be converted into the same low-dimensional space by:
\begin{equation}
  E_j = V_jx_j
\end{equation}
where $V_j \in \mathbb{R}^k$ is the corresponding field embedding with size $k$.

Through the aforementioned method, an embedding layer is applied upon the raw feature input to compress it to a low dimensional, dense real-value vector. The result of embedding layer is a wide concatenated vector:

\begin{equation}
  E = concat(E_1, E_2, ..., E_i, ..., E_f)
\end{equation}
where $f$ denotes the number of fields, and $E_i \in \mathbb{R}^k$ denotes the embedding of one field. Although the feature lengths of instances can be various, their embeddings are of the same length $f\times k$, where $k$ is the dimension of field embedding.

\subsection{Contextual Embedding}
As discussed in Section 3.1, the contextual embedding module in ContextNet has two objectives: Firstly, ContextNet use this module  to aggregate contextual information for each feature from input instance, that is to say, feature embedding layer. Secondly, the collected contextual information for one feature is projected to the same low-dimensional space as feature embedding lies in.

 We can formulate this process as  follows:

\begin{equation}
  CE_i = \mathcal{F}_{project}(\mathcal{F}_{agg}(E_i, E; \Theta_a); \Theta_p)
\end{equation}
where  $CE_i \in \mathbb{R}^k$ denotes the contextual embedding of the $i$-th feature $E_i$ and $k$ is the dimension of field embedding, $\mathcal{F}_{agg}(E_i, E; \Theta_a)$ is the contextual information aggregation function for the $i$-th feature field which uses the embedding layer $E$ and feature embedding $E_i$ as input and $\Theta_a$ denotes the parameters of aggregation model. $\mathcal{F}_{project}(F_{agg}; \Theta_p)$ is the mapping function to project the contextual information into the same low-dimensional space with feature embedding lies in. $\Theta_p$ denotes the parameters of projection model.

To make this module more specific, We propose a  two-layer contextual embedding network (TCE) for this module in our paper. That is to say, we adapt the feed forward network as the aggregation function $\mathcal{F}_{agg}(E_i, E; \Theta_a)$ and projection function $\mathcal{F}_{project}(F_{agg}; \Theta_p)$. Notice here that TCE is just a specific solution for this module and there are other options  that deserve further exploration. However, the input of the contextual embedding module should be from embedding layer which contains original and global contextual information.


\begin{figure}
\begin{minipage}[t]{0.49\linewidth}
  \centering
  \includegraphics[width=0.8\linewidth]{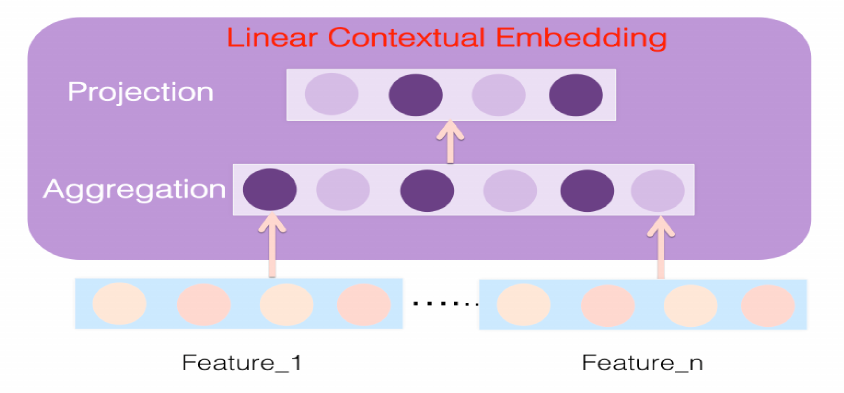}
  \caption{Two Layer  Contextual Embedding}
  \label{fig:linearcontext}
\end{minipage}%
\begin{minipage}[t]{0.49\linewidth}
  \centering
  \includegraphics[width=.8\linewidth]{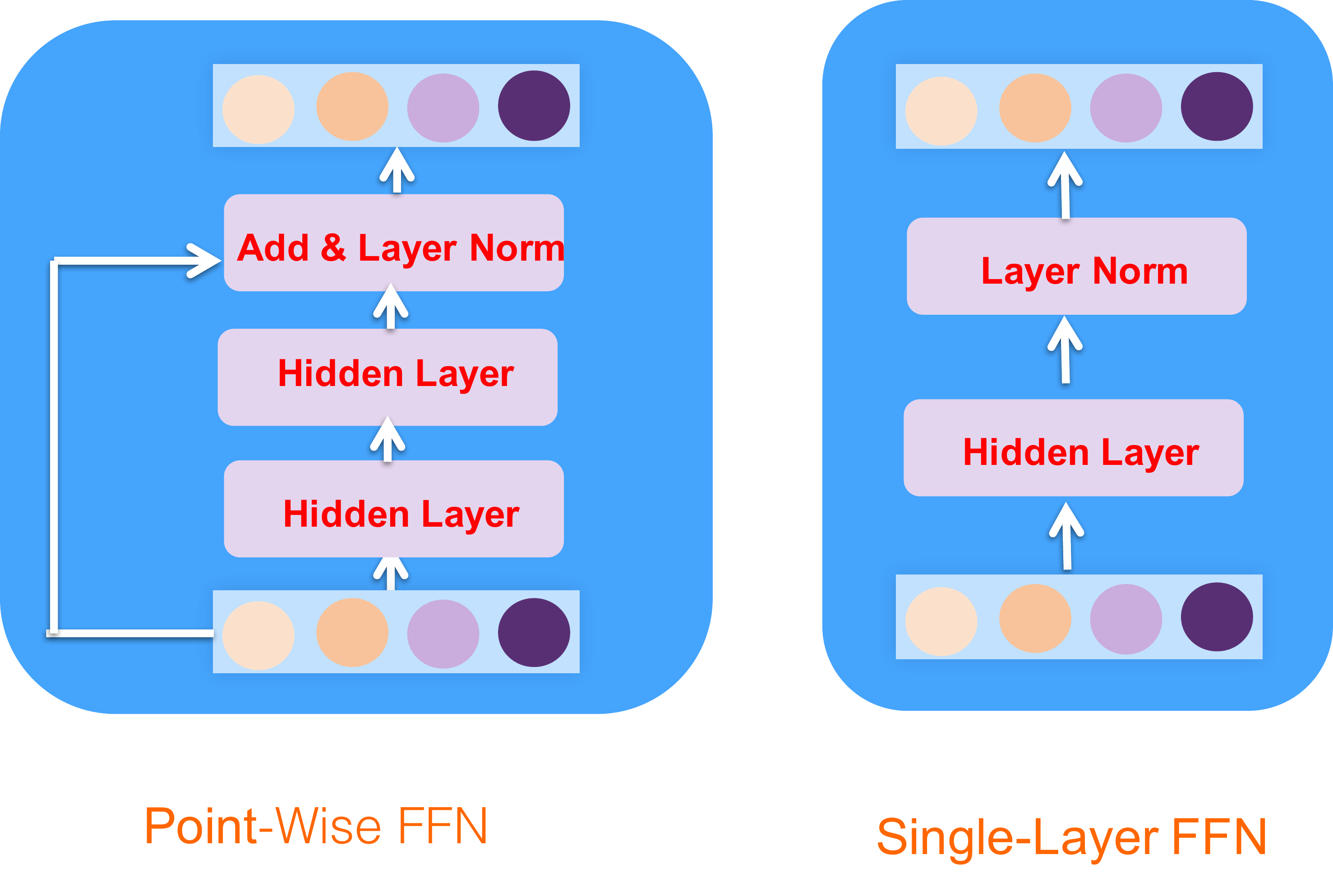}
  \caption{Structure of Non-Linear Transformation}
  \label{fig:transfer}
\end{minipage}
\end{figure}

Next we will describe how contextual embedding network works. Suppose we have a feature $E_i$ which belongs to feature field $d$. As depicted in Figure \ref{fig:linearcontext}, two fully connected (FC) layers are used in TCE module. The first FC layer is called "aggregation layer" which is a relatively wider layer to collect the contextual information from embedding layer with parameters $W^a_d$. The second FC layer is "projection layer" which projects the contextual information into the same low-dimension space with feature embedding and reduces dimensionality to the same size as feature embedding has. The projection layer has parameter $W_d^p$. Formally,
\begin{equation}
  CE_i = \mathcal{F}_{project}(\mathcal{F}_{agg}(E_i, E; \Theta_a); \Theta_p) = W_d^p(RELU(W_d^aE))
  \label{eq:second_fc_layer}
\end{equation}
where $E \in \mathbb{R}^{m=f\times k}$ refers to the embedding layer of input instance, $W_d^a \in \mathbb{R}^{t\times m}$ and $W_d^p \in \mathbb{R}^{k\times t}$ are parameters for aggregation and projection layer in TCE for field $d$, respectively. $t$ and $k$ respectively denotes the neural number of aggregation and projection layer. Notice here that the aggregation layer is usually wider than the projection layer because the size of the projection layer is required to be equal to the feature embedding size $k$. The wider aggregation layer will make the model be more  expressive.

We can see from formula (\ref{eq:second_fc_layer}) that each feature field $d$ maintains its own parameters $W_d^a$ and $W_d^p$ for aggregation and projection layer, respectively. Suppose we have $f$ different feature fields in embedding layer, the parameter number of TCE will be $f\ast(W_d^a + W_d^p)$. We can reduce parameter number by sharing $W_d^a$ in aggregation layer among all fields. The parameter number of TCE will be reduce to $W_d^a + f\ast W_d^p$. In order to reduce the model parameters, we adopt the following strategy: we share the parameters of aggregation layer among all feature fields while keep the parameters of projection layer private for each feature field. This strategy effectively balances the model complexity and  the model's expressive ability because the private projection layer will make each feature extract useful contextual information for its purpose independently. This makes the TCE module look like the "share-bottom" structure in multi-task learning where the bottom hidden layers are shared across tasks as works \cite{caruana1997multitask} and \cite{caruana1993multitask} did.

\subsection{ContextNet Block}
As discussed in Section 3.1,  ContextNet block is used to dynamically refine each feature's embedding by merging the contextual embedding produced for that feature to implicitly capture the high-order feature interactions. To achieve this goal, there are two consequential procedures in  ContenxtNet block as shown in Figure \ref{fig:contextnet}: embedding merging and a following non-linear transformation. We can stack ContextNet block by block to form deep network and the output of the previous block is the input of the next block.

Next we will describe how ContextNet block works. We use  $E_i^l$ to denote the output feature embedding of the $l$-th block, that is to say, $E_i^l$ is the input embedding of the $i$-th feature for the $(l+1)$-th block. $CE_i^{l+1}$ denotes the corresponding contextual embedding computed by TCE for the $i$-th feature field in the $(l+1)$-th block.

We can describe this process for the $i$-th feature as follows:
\begin{equation}
  E_{i}^{l+1} = \mathcal{F}_{non-linear}(\mathcal{F}_{merge}(E_{i}^{l}, CE_{i}^{l+1}; \Theta_m); \Theta_n)
\end{equation}
where $E_i^{l+1} \in \mathbb{R}^k$ denotes the fine-tuned feature embedding outputted by the $(l+1)$-th ContextNet block for the $i$-th feature $E^l_i$ and $k$ is the dimension of field embedding, $\mathcal{F}_{merge}(E_{i}^{l}, CE_{i}^{l+1}; \Theta_m)$ is the merging function for the $i$-th feature which uses the previous block's output feature embedding $E_i^{l}$ and contextual embedding $CE_i^{l+1}$ in current block as input. $\Theta_m$ denotes the parameters pf merging function. $\mathcal{F}_{non-linear}(F_{merge}; \Theta_n)$ is the mapping function to conduct the non-linear transformation on the merged embedding in order to further capture high-order interactions for the $i$-th feature. $\Theta_n$ denotes the parameters of non-linear transformation function.

As for the merging function $F_{merge}(E_i^l, CE^{l+1}_i; \Theta_m)$, Hadamard product is used  in this work to merge the feature embedding $E_i^l$ and the corresponding contextual embedding $CE_i^{l+1}$ as follows:
\begin{equation}
  E_i^{l} \otimes CE_{i}^{l+1} = \left[E_{i1}^l\cdot CE_{i1}^{l+1}, ..., E_{ij}^l\cdot CE_{ij}^{l+1}, ..., E_{ik}^l\cdot CE_{ik}^{l+1}\right]
\end{equation}

\noindent where $k$ is the size of embedding vector $E_i^l$ and contextual embedding $CE_i^{l+1}$ for the $i$-th feature. Hadamard product is a kind of element-wise production operation without parameters.


As for the non-linear function $F_{non-linear}(F_{merge}; \Theta_n)$, we propose two neural networks which are shown in Figure \ref{fig:transfer} in this paper: point-wise feed-forward network and single-layer feed-forward network.

\noindent\textbf{Point-Wise Feed-Forward Network: }

Though the ContextNet uses the contextual embedding module to aggregate all other feature's embedding and then project them to a fixed embedding size, ultimately it is still a linear model. For endowing the model with nonlinearity in order to capture high-order interactions better, we can apply a point-wise two-layer feed-forward network to all $E_i$ identically (sharing parameters among all feature field):
\begin{equation}
  \mathcal{F}_i = PFFN(E_i; \Theta_n) = LN(RELU(E_iW^1)W^2 + E_i)
\end{equation}
where $W^1$, $W^2$ are $k \times k$ matrices and $k$ is the dimension of field embedding. We also adapt the residual connection and layer normalization ($LN$) \cite{ba2016layer}. The extra  parameter number  introduced by FFN is $W^1 + W^2$ which is not large because of the parameter sharing in FFN. ContextNet with this version FFN is called "ContextNet-PFFN" in the following part of this paper.

\noindent\textbf{Single-Layer Feed-Forward Network:}

We propose another much simpler one-layer feed-forward network by reducing the residual connection and ReLU nonlinearity. We can apply this transformation to all $E_i$ identically with sharing parameters as follows:

\begin{equation}
  \mathcal{F}_i = FFN(E_i; \Theta_n) = LN(E_iW^1)
\end{equation}

\noindent where $W^1$ is $k \times k$ matrix and the bias is also discarded. The layer normalization will bring the nonlinearity to high-order feature interactions though the ReLU is removed from the mapping function. The extra  parameter number introduced by FFN is $W^1$ which is small because of the parameter sharing in FFN. We call ContextNet with this version FFN "ContextNet-SFFN" in the following part of this paper. Though much simpler in the mapping form compared with ContextNet-PFFN,  ContextNet-SFFN has comparable or even better performance in many datasets and we will discuss this in detail in Section 4.2.

The different ContextNet block has the corresponding contextual embedding module to refine each feature's embedding when we stack multi-block to form deeper network. We can further reduce the model parameter by sharing the parameters of aggregation layer or projection layer among TCE modules for each ContextNet block. The experiments are conducted about these three different parameter sharing strategies and we will discuss this in detail in Section 4.3.

\subsection{Prediction Layer}
To summarize, we give the overall formulation of our proposed model's output as:
  \begin{equation}
    \hat{y} = \delta(w_0 + \sum\nolimits_{i=1}^{f\ast k}w_ix_i)
  \end{equation}

\noindent where $\hat{y} \in (0, 1)$ is the predicted value of CTR, $\delta$ is the sigmoid function, $f$ is the feature filed number, $k$ is the feature embedding size, $x_i$ is the bit value of all feature's embedding vectors outputted by the last ContextNet block and $w_i$ is the learned weight for each bit value.

For binary classifications, the loss function is the log loss:
\begin{equation}
  \mathcal{L} = -\frac{1}{N}\sum^N_{i=1}y_i\log(\hat{y}_i)+(1-y_i)\log(1-\hat{y}_i)
\end{equation}
where $N$ is the total number of training instances, $y_i$ is the ground truth of $i$-th instance and $\hat{y}_i$ is the predicted CTR. The optimization process is to minimize the following objective function:
\begin{equation}
\mathfrak{L} = \mathcal{L} + \lambda \|\Theta\|
\end{equation}
where $\lambda$ denotes the regularization term and $\Theta$ denotes the set of parameters.

\subsection{Interpretability of  the ContextNet}
Compared with simple models such as LR\cite{10.1145/2487575.2488200}, DNN models are notorious for lack of interpretability because of the non-linearity introduced by widely used MLP layers. One advantage of ContextNet over most DNN models is that it has good model interpretability.

The last ContextBlock's outputs maintain each feature’s embedding which have merged useful high-order information and the prediction layer of ContextNet is actually a LR model. So we can easily compute each feature's weight and contribution to the final prediction score given an input instance as follows:
\begin{equation}
  FW_j = \sum_{i=1}^{k}w_iE_{ji}
\end{equation}
where $FW_j \in \mathbb{R}$ is the weight score of the $j$-th feature in an instance. $E_j \in \mathbb{R}^k$ is the embedding of the $j$-th feature outputted by last ContextBlock and $w_i$ is the learned weight in prediction layer for bit $E_{ji}$. $k$ is the size of feature embedding. Positive score leads to positive label and negative score leads to negative label.

For a specific instance, we can detect important features which can explain why the instance has the final prediction label according to formula (14). We can also compute the feature importance on the whole training set level by accumulating or averaging scores as follows:
\begin{equation}
  FW_j = \sum_{i=1}^m|FW_j^{i}|
\end{equation}

\begin{equation}
  FW_j = (\sum^m_{i=1}|FW_j^i|)/(n+\alpha)
\end{equation}

where $FW_j \in\mathbb{R}$ is the weight score of the  $j$-th feature in the whole set. $FM_j^i$ is the score for the $j$-th feature in instance $i$ and the size of the training set is $m$. The absolute value is adopted here because the score is either positive or negative. $n$ is the size of instance set where a feature appears and $\alpha$ is a norm number to reduce the influence of low frequent features. We can find out the important features according to formula (16) or discarding unimportant features with small scores  to compress model according to formula (15).

\section{Experimental Result}
We evaluate the proposed approaches on four real-world datasets and answer the following research questions:

\begin{itemize}
\item\noindent\textbf{RQ1} Does the proposed method performs better than existing state-of-the-art  deep learning based CTR models?

\item\noindent\textbf{RQ2} What is the training efficiency of ContextNet?

\item\noindent\textbf{RQ3} What is the influence of various components in the ContextNet architecture?

\item\noindent\textbf{RQ4} How does the hyper-parameters of networks influence the performance of ContextNet?

\item\noindent\textbf{RQ5} Can ContextNet really gradually refine the feature embedding to capture feature interactions? How about the feature importance computation?
\end{itemize}

In the following, we will first describe the experimental settings, followed by answering the above research questions.

\subsection{Experiment Setup}

\subsubsection{Datasets}

The following four data sets are used in our experiments:

\begin{table}
\centering
\caption{Statistics of the evaluation datasets}
\begin{tabular}{lccc}
\toprule
Datasets  & \#Instances & \#fields & \#features \\
\midrule
Criteo       & 45M  & 39 & 30M     \\
Movielens & 1M & 7 & 7478 \\
Malware     & 8.92M  & 82 & 0.97M \\
Avazu       & 40.43M  & 24 & 9.5M     \\
\bottomrule
\end{tabular}
\label{tab:datasets}
\end{table}

\begin{table*}
\centering
\caption{Overall performance (AUC) of different models on four datasets(CNet-PFFN  means ContextNet-PFFN while CNet-SFFN means ContextNet-SFFN)}
\begin{tabular}{l|cccccccc}
\toprule
  & \multicolumn{2}{c}{\textbf{Criteo}} & \multicolumn{2}{c}{\textbf{ML-1m}} &
  \multicolumn{2}{c}{\textbf{Malware}} & \multicolumn{2}{c}{\textbf{Avazu}} \\
\midrule
  & AUC & RelaImp   & AUC & RelaImp   & AUC & RelaImp   & AUC & RelaImp \\
\midrule
FM & 0.7895 & +0.00\% & 0.8446 & +0.00\% & 0.7166 & +0.00\% & 0.7785 & +0.00\%\\
DNN & 0.8054 & +5.49\% & 0.8527 & +2.35\% & 0.7246 & +3.70\% & 0.7820 & +1.26\%\\
DeepFM & 0.8057 & +5.60\% & 0.8537 & +2.64\% & 0.7293 & +5.86\% & 0.7833 & +1.72\%\\
\midrule
DCN & 0.8058 & +5.63\% & 0.8595 & +4.32\% & 0.7300 & +6.19\% & 0.7830 & +1.62\%\\
xDeepFM & 0.8064 & +5.84\% & 0.8561 & +3.34\% & 0.7310 & +6.65\% & 0.7841 & +2.01\%\\
Transformer &	0.8037 &	+4.90\% &	0.8578	& +3.83\% &	0.7267 &	+4.66\% &	0.7819 &	+1.125\% \\
AutoInt & 0.8051 & +5.39\% & 0.8569 & +3.57\% & 0.7282 & +5.36\% & 0.7824 & +1.40\%\\
\midrule
CNet-PFFN & 0.8104 & +7.22\% & 0.8641 & +5.66\% & 0.7399 & +10.76\% & 0.7862 & +2.76\%\\
CNet-SFFN & \textbf{0.8107} & \textbf{+7.32\%} & \textbf{0.8681} & \textbf{+6.82\%} & \textbf{0.7408} & \textbf{+11.17\%} & \textbf{0.7863} & \textbf{+2.80\%} \\
\bottomrule
\end{tabular}
\label{tab:overalperformance}
\end{table*}

\begin{enumerate}
  \item \textbf{Criteo\footnote{Criteo \url{http://labs.criteo.com/downloads/download-terabyte-click-logs/}} Dataset:}
  As a very famous public real world display ad dataset with each ad display information and corresponding user click feedback, Criteo data set is widely used in many CTR model evaluation.  There are $26$ anonymous categorical fields and $13$ continuous feature fields in Criteo data set.

  \item \textbf{MovieLens\footnote{MovieLens 1m.   https://grouplens.org/datasets/movielens/1m/} Dataset:}
  MovieLens is a popular benchmark dataset for evaluating recommendation algorithms. We adopt the well-established MovieLens 1m(ML-1m) as our evaluation dataset in this work, which contains $1$ million ratings from $6000$ users on $4000$ movies.

  \item \textbf{Malware \footnote{Malware https://www.kaggle.com/c/microsoft-malware-prediction} Dataset:}
Malware is a dataset to predict a Windows machine's probability of getting infected. The malware prediction task can be formulated as a binary classification problem like a typical CTR estimation task does.

  \item \textbf{Avazu\footnote{Avazu http://www.kaggle.com/c/avazu-ctr-prediction} Dataset:}
    The Avazu dataset consists of several days of ad click-through data which is ordered chronologically. For each click data, there are $24$ fields which indicate elements of a single ad impression.
\end{enumerate}

We randomly split instances by 8:1:1 for training , validation and test while Table \ref{tab:datasets} lists the statistics of the evaluation datasets.

\subsubsection{Evaluation Metrics}

AUC (Area Under ROC) is used in our experiments as the evaluation metric. AUC's upper bound is 1 and larger value indicates a better performance.

RelaImp is also adopted as work \cite{inproceedings} does to measure the relative AUC improvements over the corresponding baseline model as another evaluation metric. Since AUC is $0.5$ from a random strategy, we can remove the constant part of the AUC score and formalize the RelaImp as:
\begin{equation}
  RelaImp = \frac{AUC(Measured\ Model) - 0.5}{AUC(Base\ Model) - 0.5} - 1
\end{equation}

Log loss is another widely used metric in binary classification, measuring the distance between two distributions. The log loss results of our experiments show similar trends with AUC, so we didn't  present performances in this metric because of the limited space of the paper.

\subsubsection{Models for Comparisons}
We compare the performance of the following models with our proposed approaches: FM, DNN, DeepFM, Deep\&Cross Network(DCN), xDeepFM, Transformer and AutoInt Model, all of which are discussed in Section 2. For DCN, xDeepFM, Transformer and AutoInt, $3$-order feature interaction network structure is adopted as default setting as the original papers use. FM is considered as the base model in evaluation.

\subsubsection{Implementation Details}
We implement all the models with Tensorflow in our experiments. For optimization method, we use the Adam with a mini-batch size of $1024$ and a learning rate is set to $0.0001$.  Focusing on neural networks structures in our paper, we make the dimension of field embedding for all models to be a fixed value of $10$. For models with DNN part, the depth of hidden layers is set to $3$, the number of neurons per layer is $400$, all activation function are ReLU. Except for special mention, we have the default settings as follows for ContextNet: feature embedding is 10, default model has 3 ContextBlocks and hidden layer size of TCE module is 20. We conduct our experiments with $2$ Tesla $K40$ GPUs.

\subsection{Performance Comparison (RQ1)}
The overall performance for CTR prediction of different models on four evaluation datasets is shown in Table \ref{tab:overalperformance}. We have the following key observations:

\begin{enumerate}
  \item ContextNet achieves the best performance on all four datasets and obtains significant improvements over the state-of-the-art methods. It can boost the accuracy over the baseline FM by $2.80\%$ to $11.17\%$, baseline DeepFM by $1.06\%$ to $5.02\%$, as well as the best of DNN baselines by $0.78\%$ to $4.24\%$. We also conduct a significance test to verify that our proposed models outperforms baselines with the significance level $\alpha = 0.01$. It explicitly proves the proposed ContextNet  indeed yields strong learning capacity by modeling high-order interactions implicitly using the feature contextual embedding.

  \item 	As for the comparison of ContextNet-PFFN and ContextNet-SFFN, we can see from Table \ref{tab:overalperformance} that ContextNet-SFFN consistently outperforms ContextNet-PFFN model on all four datasets with the same settings, though it's much simpler in model structure and has less parameters. This means ContextNet-SFFN is a more applicable model in real-world applications.

  \item For models which explicitly introduce high-order feature interactions by sub-network such as DCN, xDeepFM , Transformer and AutoInt, xDeepFM outperforms other two models on three datasets while DCN's performance is best on ML-1m dataset. Compared with DCN and xDeepFM, Both AutoInt and Transformer model show no advantage on any dataset. That means feature interaction in multiplicative way after feature aggregation by a specific network indeed has an advantage over feature aggregation in summation way after pair-wise interaction as AutoInt and Transformer did.
\end{enumerate}

\subsection{Model Efficiency (RQ2)}
As mentioned in Section 3.4, In order to reduce the parameters of the linear contextual embedding module, we can share the parameters in TCE module for different ContextNet block. We have the following three strategies: 1) share the parameters in aggregation layer among corresponding TCE modules for each block(Share-A); 2) share the parameters in both aggregation layer and projection layer(Share-A\&P); 3) we don't share parameters(Share-Nothing), which is a default setting for all other experiments if not specially mentioned.

We conduct some experiments to explore the influence of the different parameter-sharing strategies on the model performance and Table \ref{tab:paremeter-share} shows the results. We can see from the results that the performances of Share-Nothing and Share-A strategies are comparable on two datasets. However, the performance will degrade greatly if we share the parameters both in aggregation layer and projection layer. This indicates that it's very critical for the model's good performance for TCE module to extract different high-order interaction information for each refined feature of  different ContextNet block. Compared with Share-Nothing strategy, Share-A is maybe a better choice because it has less parameter and can maintain good model performance.

\begin{table}
\centering
\caption{ Overall performance (AUC) of different parameter sharing strategies of  Linear Contextual Module in ContextNet-PFFN)}
\begin{tabular}{l|cc}
\toprule
  & \textbf{Criteo} & \textbf{Malware}   \\
\midrule
Share-Nothing & 0.8104 & 0.7399 \\
Share-A &  0.8094 &  0.7400 \\
Share-A\&P & 0.7926 & 0.7117 \\
\bottomrule
\end{tabular}
\label{tab:paremeter-share}
\end{table}



\begin{figure}
  \begin{minipage}[b]{.45\linewidth}
    \centering
    \captionof{table}{Overall performance (AUC) of models removing different components of ContextNet-PFFN)}
    \begin{tabular}{l|cc}
    \toprule
      & \textbf{Criteo} & \textbf{Malware}   \\
    \midrule
    ContextNet-PFFN & 0.8104 & 0.7399 \\
    -w/o TCE &  0.7923 &  0.7125 \\
      -w/o FFN & 0.8043 & 0.7354 \\
      -w/o LN & 0.8069 & 0.7373 \\
      -w/o RC & 0.8098 & 0.7380 \\
    \bottomrule
    \end{tabular}
    \label{tab:contextnet-fppn}
    \end{minipage}
    \begin{minipage}[b]{.45\linewidth}
      \centering
      \includegraphics[width=6cm]{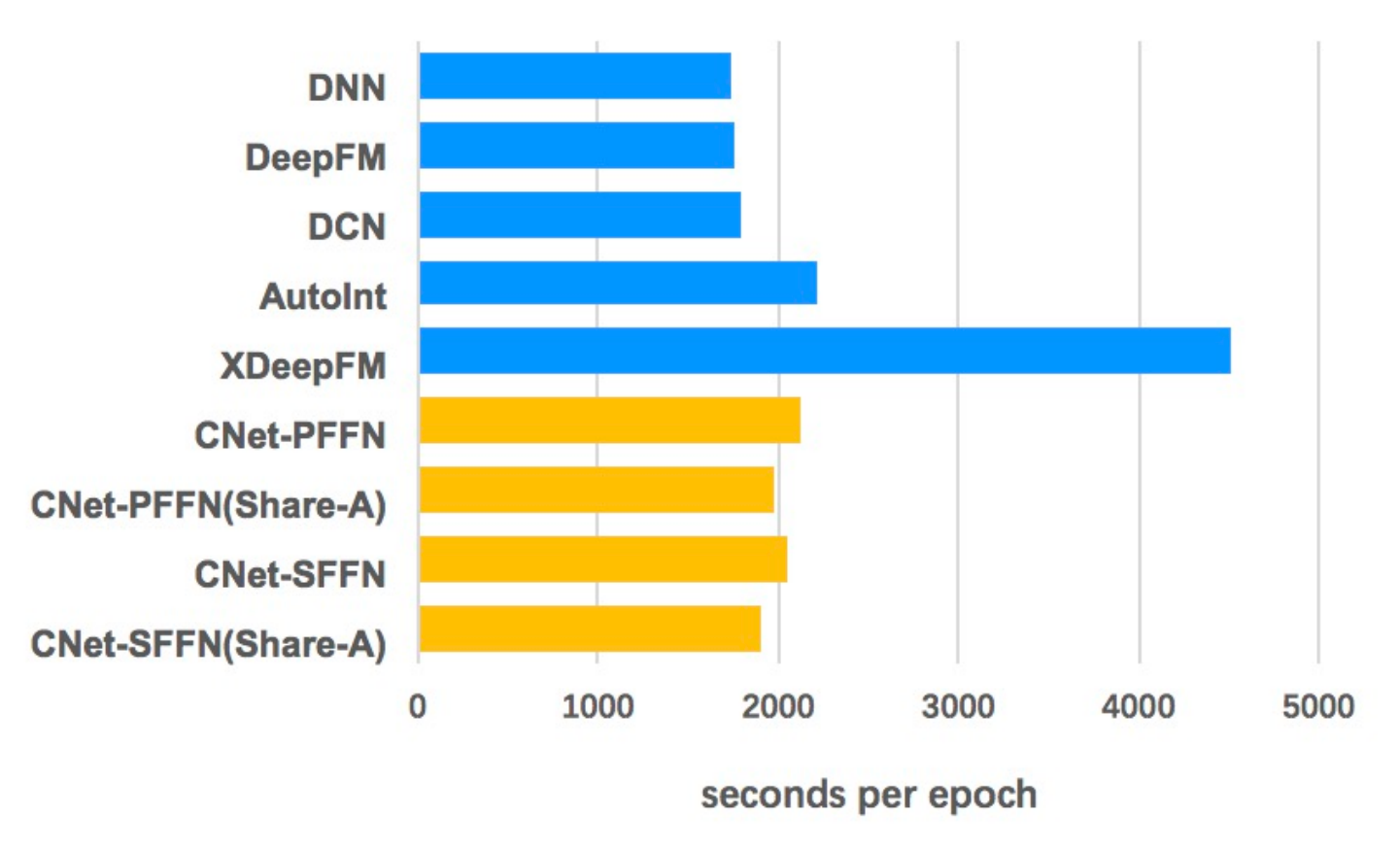}
      \caption{Efficiency comparison of different models in terms of run time per epoch on Criteo dataset}
      \label{fig:efficiency_cp}
    \end{minipage}%
\end{figure}

To compare the model efficiency of different models, we use the runtime per epoch as evaluation metric. DNN and DeepFM are regarded as efficiency baseline because they are relatively simple in network structure and are widely used in many real life applications. The comparison is conducted on Criteo dataset and the results are shown in Figure \ref{fig:efficiency_cp}. xDeepFM is much more time-consuming compared with all the other models and this implies xDeepFM is hard to be applied in many real life scenarios. As for the training efficiency of our proposed ContextNet models, we can see that both the ContextNet-PFFN and ContextNet-SFFN have faster training speed compared with AutoInt and xDeepFM. If we share the parameters in aggregation layer of two proposed ContextNet models, the training speed can be further increased. The ContextNet-SFFN model with Share-A strategy can run just slightly slower than baseline model, which means that ContextNet is sufficiently efficient for real world applications.


\subsection{Ablation Study (RQ3)}
In this section, we perform ablation experiments over key components of ContextNet in order to better understand their impacts on Criteo and Malware datasets(the other two datasets show similar trend), including linear contextual embedding (TCE), feed-forward network (FFN), layer normalization (LN), and residual connection (RC) . Table \ref{tab:contextnet-fppn} shows the results of our default version (Block = $3$), and its variants on two datasets.


\begin{enumerate}
  \item Remove TCE: The performance of ContextNet-PFFN dramatically degrades on both datasets without TCE module. It tells us that the contextual information gathered by TCE module is critical for the ContextNet and we deem TCE module extracts different high-order interactions information for various features.

  \item Remove FFN: Without FFN, the model performance also degrades obviously and that may indicate the non-linear transformation on the result of element-wise product of feature embedding and contextual information is also important for ContextNet.

  \item Remove LN or RC. From the results in Table \ref{tab:contextnet-fppn}, we can see that removing either LN or RC also decreases model performance, though the performance degradation is not as much as that of  removing ICE and FFN.
\end{enumerate}

We can see the usefulness of the different components of ContextNet from the above-mentioned ablation studies.

\begin{figure}
  \setlength{\abovecaptionskip}{0pt}
  \centering
  \includegraphics[width=0.69\linewidth,height=3.5cm]{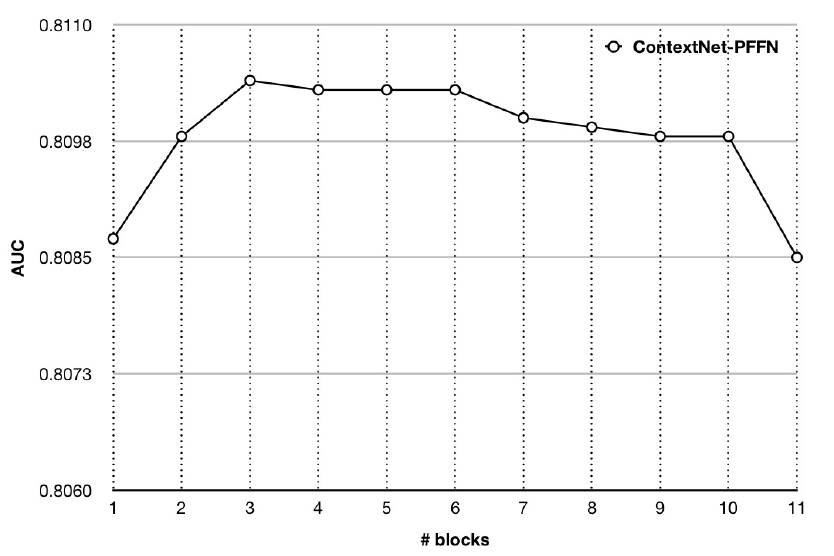}
  \caption{Effect of Different Blocks on Model Performance}
  \label{fig:blocks}
\end{figure}

\subsection{Hyper-Parameter Study(RQ4)}

In this section, we study the impact of hyper-parameters on ContextNet, including (1) the number of ContextNet blocks; (2) the number of feature embedding size. The experiments are conducted on Criteo and Malware datasets via changing one hyper-parameter while holding the other settings. Other two datasets show the similar trends and we didn't present them because of the limited space.

\paragraph{\textbf{Number of ContextNet Blocks.}}   To explore the influence of the number of ContextNet's blocks on model's performance, we conduct some experiments on Criteo dataset to stack blocks of ContextNet-PFFN model from $1$ block to $11$ blocks. Figure \ref{fig:blocks} shows the experimental results. It can be observe that the performance increases with more blocks at the beginning and  the performance can be maintained until the number is set greater than $10$. Considering there are $4$ hidden layers in one block of Context-PFFN model, we can see that ContextNet-PFFN has the depth of nearly $40$ hidden layers in the model while keeping good performance. This may indicate that contextual embedding module helps the trainability of very deep network in CTR tasks.

\paragraph{\textbf{Number of Feature Embedding Size.}} The results in Table \ref{tab:differentembeddingsize} show the impact of the number of feature embedding size on model performance. We can observe that the performance of ContextNet-PFFN increases with the increase of embedding size at the beginning. However, model performance degrades when the embedding size is set greater than $50$. Over-fitting of deep network is maybe the reason for this. Compared with the performance of  the model  shown in Table \ref{tab:overalperformance} which has a embedding size of $10$, the bigger embedding size further increases model's performance on two datasets.

\begin{table}[h]
\centering
\caption{Overall performance (AUC) of different feature embedding size of ContextNet-PFFN}
\begin{tabular}{l|ccccc}
\toprule
 & \textbf{10} & \textbf{20} & \textbf{30} & \textbf{50} & \textbf{80}  \\
\midrule
   Criteo & 0.8104 & 0.8109 & 0.8111 & 0.8113 & 0.8097 \\
   Malware & 0.7399 & 0.7416 & 0.7417 & 0.7414 & 0.7405 \\
\bottomrule
\end{tabular}
\label{tab:differentembeddingsize}
\end{table}

\begin{figure}
  \setlength{\abovecaptionskip}{0pt}
  \centering
  \includegraphics[width=.89\linewidth]{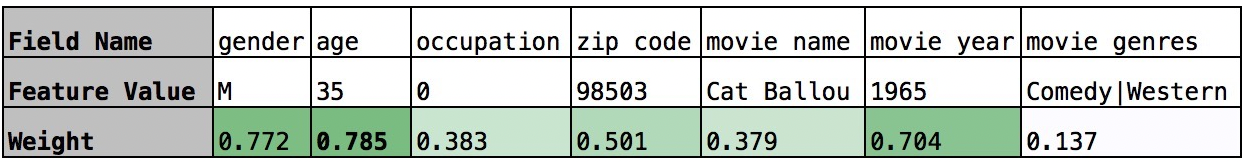}
  \caption{Example Instance from ML-1m}
  \label{fig.fig5}
\end{figure}

\subsection{Analysis of Dynamic Feature Embedding (RQ5)}

\begin{figure}
  \setlength{\abovecaptionskip}{2pt}
  \centering
  \includegraphics[width=.25\linewidth]{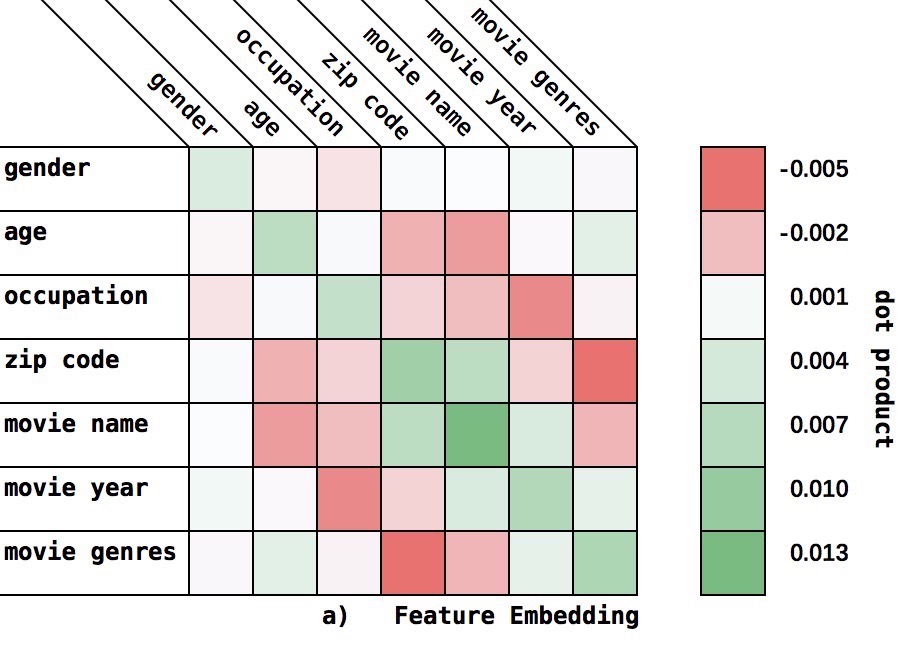}\hfill
  \includegraphics[width=.25\linewidth]{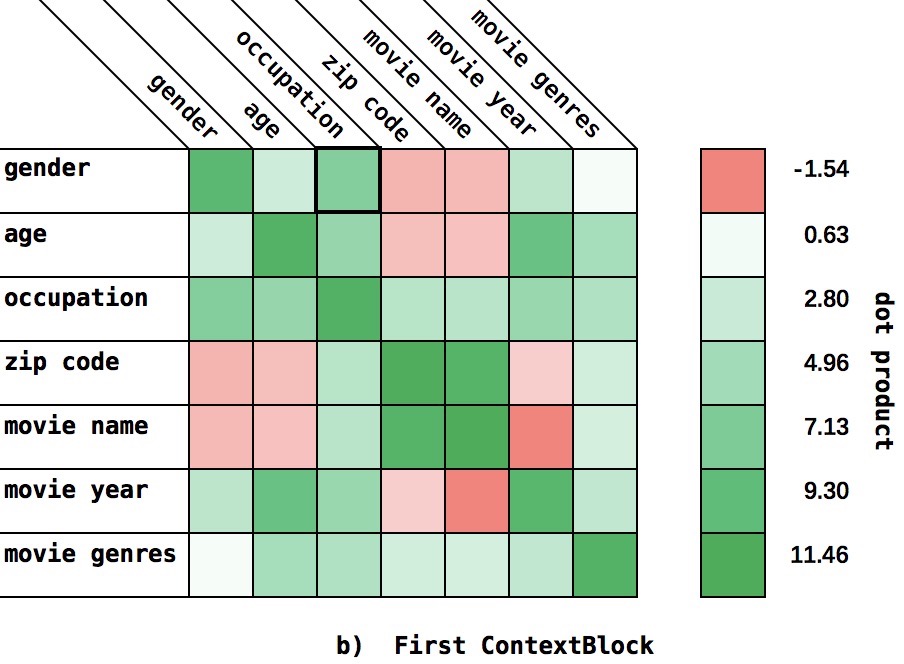}
  \includegraphics[width=.25\linewidth]{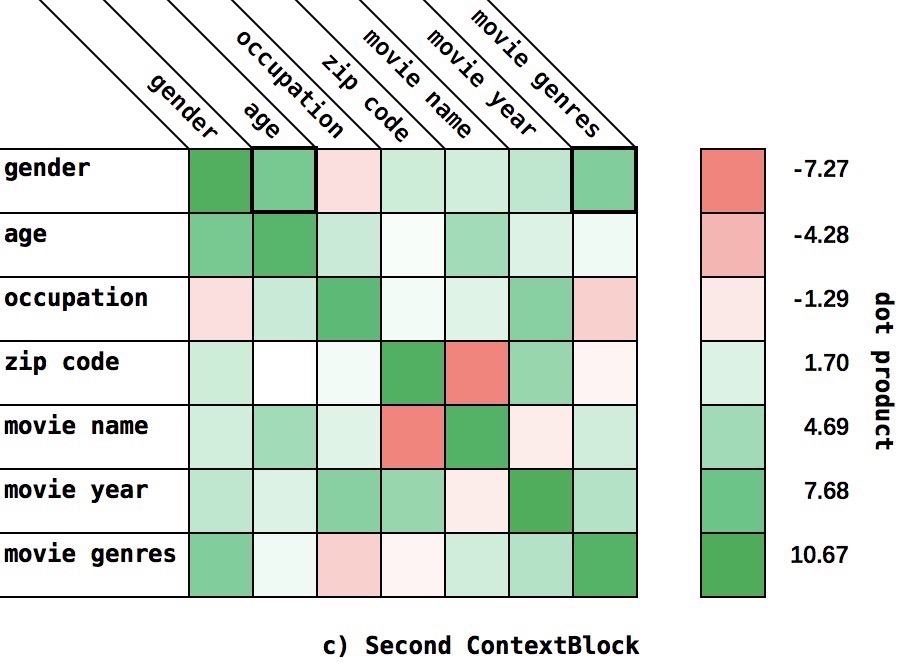}\hfill
  \includegraphics[width=.25\linewidth]{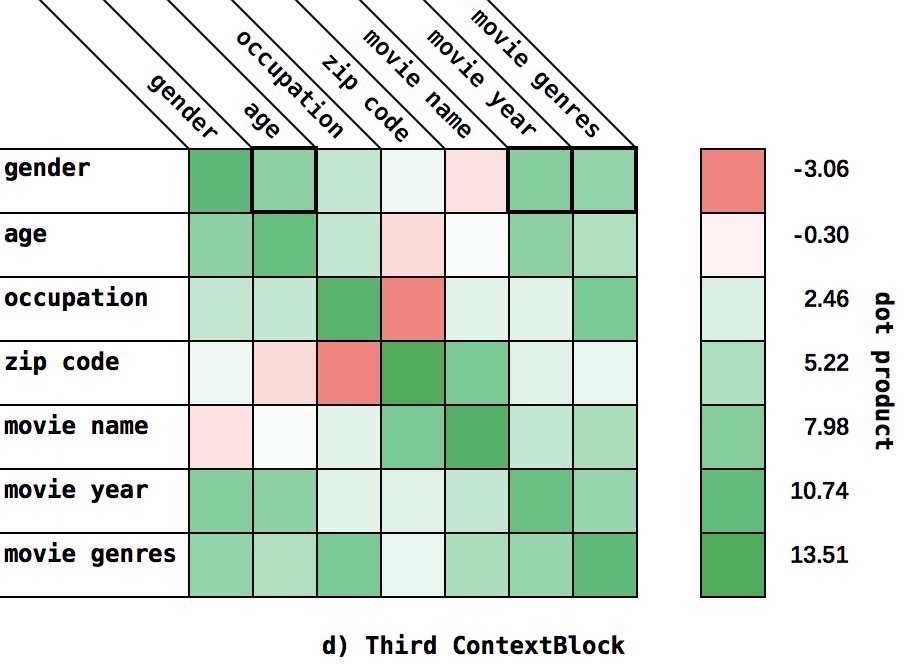}
  \label{fig.featureemb}
  \caption{Analysis of Dynamic Feature Embedding}
\end{figure}

\begin{figure}
  \setlength{\abovecaptionskip}{2pt}
  \centering
  \includegraphics[width=.45\linewidth]{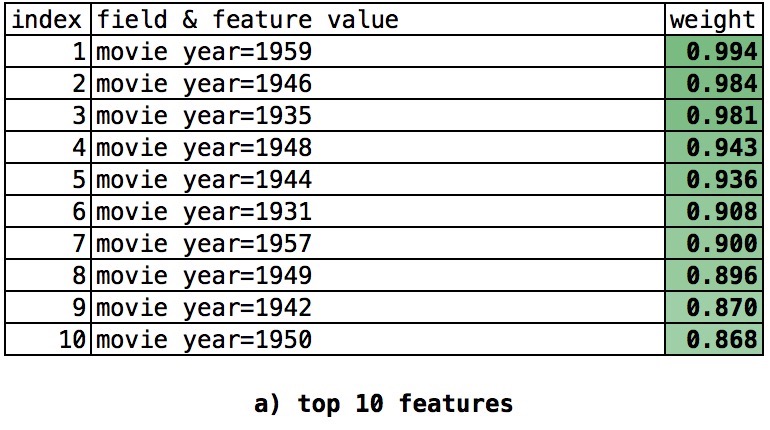}\hfill
  \includegraphics[width=.45\linewidth]{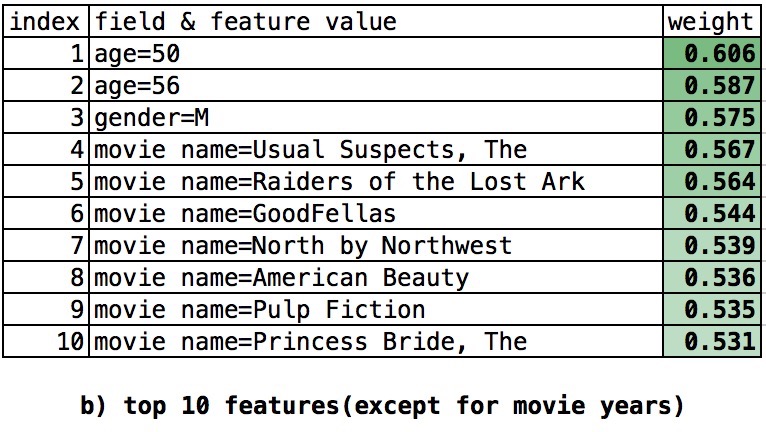}
  \label{fig.fig8}
  \caption{Top 10 Features of ML-1m Dataset}
\end{figure}

To verify that ContextNet indeed dynamically changes each feature’s embedding block by block to capture feature interactions, we input a randomly sampled instance (Figure \ref{fig.fig5}, the instance has positive label with estimated CTR score $0.975$)  from ML-1m dataset into trained ContextNet-PFFN. Then we compute the dot product for each feature pair using each ContextBlock’s outputted feature embedding, including the feature embedding layer. The high positive dot product value means the two feature have similar embedding content and are highly correlated interactions.  The Figure 7 shows the result. The deeper green color means bigger positive value while deeper red color denotes bigger negative dot product value. The elements on the diagonal of the matrix can be ignored because that's  each feature's  self dot product.  We can see from the Figure 7 that:

For features in feature embedding layer, most dot product values are small numbers near zero, which indicates the embedding values are very small and there is no correlation between different fields of input features. ( Fig. a of Figure 7). However, each feature’s embedding is dynamically changed by ContextBlock to gradually find the most useful feature interactions for its own purpose: more bigger dot product values begin to appear which means correlations between features. Take the field "gender" as an example, if we adopt $5.0$ as the threshold, the highly correlated features change from "occupation" ( Fig. b of Figure 7) to "age" and "movie genres" ( Fig. c of Figure 7). The final useful feature interactions focus on the features from field
"age", "movie year" and "movie genres" ( Fig. d of Figure 7).

As for the feature importance analysis, we provide examples shown in Figure \ref{fig.fig5} and Figure 8. The feature contributes most to the final prediction score is {"age"=35} in this instance(Figure \ref{fig.fig5}). Figure 8 shows the most important features  in ML-1m dataset according to formula (16) in Section 3.6.

\section{Conclusion}

In this paper, We firstly propose a novel CTR Framework named ContextNet that implicitly models high-order feature interactions by dynamically refining the feature embedding. We also propose two specific models(ContextNet-PFFN and ContextNet-SFFN) under this framework. We conduct extensive experiments on four real-world datasets and the experiment results demonstrate that our proposed ContextNet-PFFN and ContextNet-SFFN model outperform state-of-the-art models such as DeepFM and xDeepFM significantly.

\bibliographystyle{ACM-Reference-Format}
\bibliography{context}

\end{document}